\documentclass{article}

\usepackage{blindtext} 

\usepackage[sc]{mathpazo} 
\usepackage[T1]{fontenc} 
\usepackage{microtype} 

\usepackage[english]{babel} 
\usepackage{float}
\usepackage[hmarginratio=1:1,top=32mm,columnsep=20pt]{geometry} 
\usepackage[hang, small,labelfont=bf,up,textfont=it,up]{caption} 
\usepackage{booktabs} 
\usepackage{array} 
\usepackage{paralist} 
\usepackage{verbatim} 
\usepackage{subfig} 
\usepackage{lettrine} 
\usepackage{graphicx} 
\usepackage{wrapfig}
\usepackage{multicol}
\usepackage{mathtools}
\usepackage{bbm}
\usepackage{textgreek}
\usepackage[rightcaption]{sidecap}
\usepackage{setspace}
\usepackage{eufrak}
\usepackage{stix2}
\usepackage{amsthm}

\newtheorem{theorem}{Theorem}

\newtheorem{proposition}{Proposition}

\newtheorem{lemma}{Lemma}

\usepackage{enumitem} 
\setlist[itemize]{noitemsep} 

\usepackage{abstract} 

\usepackage{titlesec} 
\renewcommand\thesection{\Roman{section}} 
\renewcommand\thesubsection{\roman{subsection}} 
\titleformat{\section}[block]{\large\scshape\centering}{\thesection.}{1em}{} 
\titleformat{\subsection}[block]{\large}{\thesubsection.}{1em}{} 

\usepackage{fancyhdr} 
\usepackage{titling} 

\usepackage{hyperref} 

\pretitle{\begin{center}\Huge\bfseries} 
\posttitle{\end{center}} 
\title{Syntropy in complex systems: A complement to Shannon's Entropy} 
\author{%
\textsc{Santiago Mendez-Moreno}\\
\\[1ex] 
\normalsize Universidad Autónoma de San Luis Potosí \\ 
\normalsize \href{mailto:}{}\\
\\
\normalsize  86 word abstract\\
\normalsize  13 figures\\
\normalsize No tables\\
}
\date{\today} 

\clearpage

\renewcommand{\maketitlehookd}{%

\begin{abstract}

This study introduces the syntropy function ($S_N$) and expectancy function ($E_N$), derived from the novel function \(\phi\), to provide a refined perspective on complexity, extending beyond conventional entropy analysis. $S_N$ is designed to detect localized coherent events, whereas $E_N$ encapsulates expected system behaviors, offering a comprehensive framework for understanding system dynamics. The manuscript explores essential theorems and properties, underscoring their theoretical and practical implications. Future research will further elucidate their roles, particularly in biological signals and dynamic systems, suggesting a deep interplay between order and chaos.\\

keywords: Syntropy Function;Expectancy Function;Complex Analysis;System Dynamics;Order and Chaos;Entropy\\

\textbf{Physics and Astronomy Classification Scheme (PhySH) Concepts:} Statistical physics, Information theory, Nonlinear dynamics, Biological physics, Computational physics.
\end{abstract}
}

\begin{document}

\maketitle

\section*{Authors Biography}
\textsc{Santiago Mendez-Moreno, M.C.;}\\
\normalsize{ Universidad del Valle de México, Ingeniería Mecatrónica, 2016;}\\
\normalsize{ Universidad Autónoma de San Luis Potosí, M.C., 2020.;}\\
\normalsize{ Universidad Autónoma de San Luis Potosí, Ph.D. student, 2021-now.}\\
\\
\clearpage
\section{Introduction}

The exploration of complexity in biological and physical systems remains a pivotal area of research, reflecting a spectrum of methodologies to quantify system dynamics. Among these, entropy, rooted in thermodynamics and information theory, is traditionally employed to gauge a system's disorder or unpredictability \cite{Shannon1948}. Yet, entropy's scope in encapsulating complexity is not absolute. A nuanced perspective suggests that complexity embodies a symbiotic coexistence of order and disorder, hinting at underlying structures of integration and differentiation.\\

Addressing this gap, we introduce "syntropy" along with its mathematical embodiments: the expectancy function ($E_N$), temporal syntropy, spectral syntropy and dynamic syntropy. These constructs are envisioned to augment the entropy paradigm, offering a lens to discern ordered, predictable, and integrated facets within a system. This discourse unfolds a comprehensive mathematical narrative of $S_N$, exploring their attributes and their utility in enriching our understanding of complexity. Through this, we elucidate how this function, in concert with entropy, unveil the intricate interplay of order and chaos pervasive in complex systems.\\

Through theorems, propositions, and applications, we illuminate their significance in capturing the essence of complex systems, offering a comprehensive narrative on the balance of predictability and spontaneity in nature.

\clearpage
\section{Mathematical Formulation}

\begin{lemma}
For all convex function that has a minimum in the uniform distribution and the maximums on the extremes of the simplex.
\end{lemma}

\begin{multicols}{2}
From Figure \ref{simplex}, then:\\

 Let $\phi$ be a function in $\mathbbm{R}$ and let it be convex, then:

$$C_N \sum \limits_{i=1}^{N} \phi(\alpha_i) \geq N C_N \phi(\frac{1}{N})$$

We choose $C_N = \frac{1}{N \phi(1/N)}$, such that:

$$ C_N \sum \limits_{i=1}^{N} \phi (\alpha_i) \geq 1/N$$

Now, $\phi:[0,1]\rightarrow \mathbbm{R}^+$, being convex, bounded and strictly increasing:

$$\phi(\sum \limits_{i=1}^{N} \alpha_i) \leq \sum \limits_{i=1}^{N} \phi(\alpha_i)$$

with $\sum \limits_{i=1}^{N} \alpha_i = 1$, then:

$$C_N \sum \limits_{i=1}^{N} \phi(\alpha_i) = N C_N \sum \limits_{i=1}^{N} \frac{\phi(\alpha_i)}{N} \geq N C_N \phi(\sum \limits_{i=1}^{N} \frac{\alpha_i}{N})$$

$$=N C_N \phi(\frac{1}{N})$$

Therefore:

$$E_N(\alpha) = C_N \sum \limits_{i=1}^{N} \phi(\alpha_i)$$

We choose $\phi(\alpha_i) = e^{\alpha_i} - 1$, which give us:

$$C_N = \frac{1}{N(e^{1/N}-1)}$$ thus

$$E_N(\alpha) = \frac{1}{N(e^{1/N}-1)} \sum \limits_{i=1}^{N} (e^{\alpha_i}-1)$$

\end{multicols}

The new proposed function denoted as $E_N$ should be called Expectancy. This function provides a measure of the expected value of a given variable \(\alpha\), normalized over \(N\) observations. The normalization factor ensures that the expectancy remains within a defined range, making it suitable for comparative analyses across different data sets.

\includegraphics[width=0.8\linewidth]{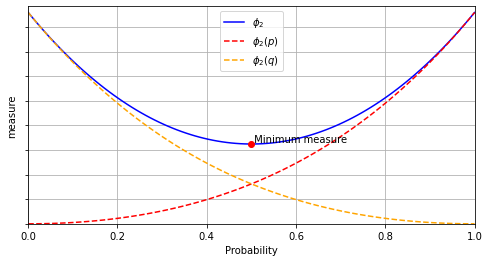}
\captionof{figure}{Simplex with minimum in $\frac{1}{N}$ and maximum in $\delta(0)$ and $\delta(1)$ for a probability '$p$' and '$q = 1-p$'.}
\label{simplex}

\clearpage
\section{Theorems and Propositions}

In this section, we will discuss the theorems and propositions of the expectancy function $E_N$ and the syntropy function $S_N$. These properties are derived from the mathematical formulation of these functions and provide insights into their behavior and potential applications.

\subsection{Theorems}
\begin{multicols}{2}
\begin{theorem}
Let $g$ be a function defined by the following relation $g(N) = \frac{1}{N(\sqrt[N]{e}-1)}$ with $N \in \mathbbm{N}^+$ , then, when $N \rightarrow \infty, g \leq M$.
\end{theorem}

\begin{proof}
$$\lim\limits_{N \to \infty} g(N) =\lim\limits_{N \to \infty} \frac{1}{N(\sqrt[N]{e}-1)}$$\\
\text{Applying L´Hopital where:}\\
$$\lim\limits_{N \to \infty} \frac{p(N)}{q(N)}=\lim\limits_{N \to \infty} \frac{p'(N)}{q'(N)}$$\\
$$\text{Make: } p(N) = \frac{1}{N} \text{ and } q(N) = \sqrt[N]{e}-1$$\\
$$\text{Then: } p'(N)=-N^{-2} \text{ and } q'(N)=-N^{-2}\sqrt[N]{e}$$\\
$$\lim\limits_{N \to \infty} \frac{p'(N)}{q'(N)}=\lim\limits_{N \to \infty}\frac{N^2}{N^2\sqrt[N]{e}}=\frac{1}{e^0}$$\\
$$\lim\limits_{N \to \infty} \frac{1}{N(\sqrt[N]{e}-1)} = 1 \text{ therefore } M=1 $$\\
\end{proof}

\begin{theorem}
The function $S_N(\textbf{A}) < M$, where $\forall \alpha \in [0,1)$ and $N \rightarrow \infty$
\end{theorem}

\begin{proof}
$$S_N(\alpha)= \frac{e^\alpha-1}{N(\sqrt[N]{e}-1)}=(e^\alpha-1)\frac{1}{N(\sqrt[N]{e}-1)}$$\\
$$\text{Let } \frac{1}{N(\sqrt[N]{e}-1)}=g(N) \text{ and } e^\alpha-1=f(\alpha)$$\\
$$\text{Then }S_N(\textbf{A})\equiv g(N)f(\alpha)$$\\
$$\text{Therefore } \lim\limits_{N \to \infty;\alpha \to 1}g(N)f(\alpha)=\lim\limits_{N \to \infty}g(N)\lim\limits_{\alpha \to 1}f(\alpha)$$\\
$$\text{From Theorem 1 we get } \lim\limits_{N \to \infty}g(N) = 1$$\\
$$\text{and finally} \lim\limits_{N \to \infty}g(N)\lim\limits_{\alpha \to 1}f(\alpha)=(e-1)$$\\
\end{proof}

\begin{theorem}
The function $E_N(\textbf{A})$ is monotonus and inyective in the interval $\mathbbm{R} [0,1)$ and for all $N \in \mathbbm{N}^+$.
\end{theorem}

\begin{figure}[H]
\centering
\includegraphics[width=0.5\linewidth]{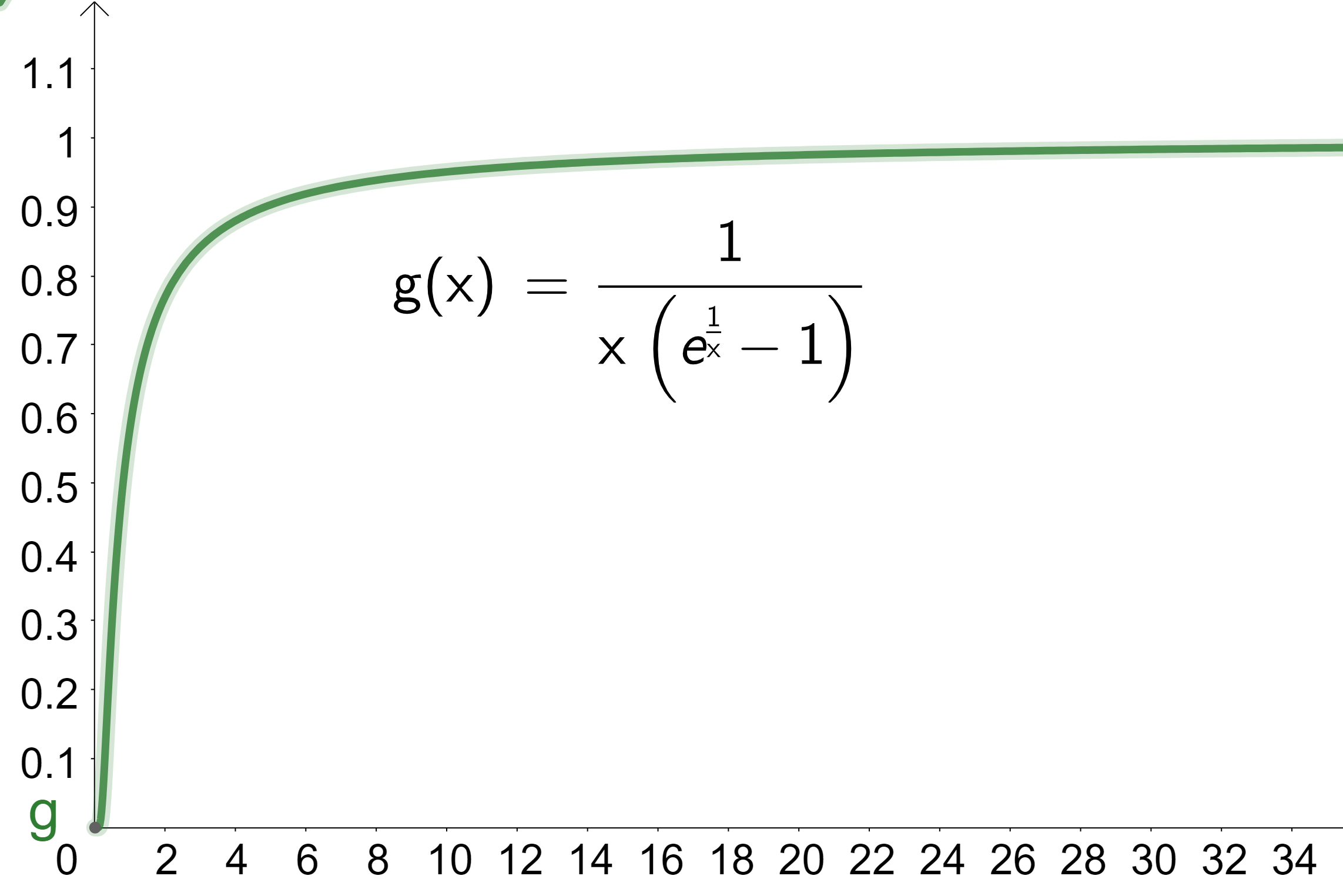}
\caption{The plot displays the behavior of the $g(N)$ function for positive values of $N$, where it approaches zero as $N$ approaches zero and asymptotically approaches 1 as 
$N$ increases.}
\label{one}
\end{figure}

\begin{proof}
Through graphic method (figure \ref{one}) we have that there is one inflection point in the range $(1,\infty)$, for $g(N)$ to be positively concave and thus monotonus and inyective, then the slope must be positive in all $(1,\infty)$, for that we have:

$$\frac{d (g(N))}{dN} = \frac{\sqrt[N]{e}-1-\frac{1}{N}\sqrt[N]{e}}{N^2 (\sqrt[N]{e}-1)^2}$$\\
$$\text{Let } N \to \infty, \lim\limits_{N \to \infty}\frac{\sqrt[N]{e}-1-\frac{1}{N}\sqrt[N]{e}}{N^2 (\sqrt[N]{e}-1)^2}$$\\
$$\text{then} \lim\limits_{N \to \infty}(\sqrt[N]{e}-1-\frac{1}{N}\sqrt[N]{e})g(N)=$$\\
$$\lim\limits_{N \to \infty}(\sqrt[N]{e}-1-\frac{1}{N}\sqrt[N]{e})\cdot \lim\limits_{N \to \infty}g(N) \cdot \lim\limits_{N \to \infty}g(N)$$\\
$$\text{From Theorem 1 we get: } \lim\limits_{N \to \infty}g(N)=1$$\\
$$\text{and } \lim\limits_{N \to \infty}\sqrt[N]{e} = e^{1/\infty} = e^0 = 1$$\\
$$\text{and } \lim\limits_{N \to \infty}\frac{1}{N} = 0$$\\
$$\therefore \lim\limits_{N \to \infty}\frac{\sqrt[N]{e}-1-\frac{1}{N}\sqrt[N]{e}}{N^2 (\sqrt[N]{e}-1)^2}=0 $$\\
\end{proof}
\end{multicols}

\clearpage
\subsection{Propositions}
\begin{multicols}{2}
\begin{proposition}
For any $\textbf{X}\triangleq \{x_i| \forall i \in N\}$,\quad with \quad $P[\textbf{X} = x_i] = \alpha_i$, \quad $\beta_N \sum \limits_{i=0}^{N-1} (e^{\alpha_i}-1)\leq M$.
\end{proposition}

Let \textbf{X} be a discrete random variable with $N$ possible values and a probability associated to each value of  $\alpha_i$. The proposition is true when $ \alpha_i=\alpha_j, \forall i , j \in \{1,2,..., N\}$\\

a) When all events have the same probability, then the expectancy is at its lowest.\\
\begin{center}
\begin{spacing}{2.5}
let $N\to \infty$ and $P[x_i] = \alpha_i =\frac{1}{N}, \forall i \in N$ then\\
$\sum \limits_{i=0}^{N-1} \beta_s (e^{\alpha_i}-1)$
$\sum \limits_{i=0}^{N-1} \beta_s (e^{\frac{1}{N}}-1) = N \beta_s (e^{\frac{1}{N}}-1) = \frac{N (e^{1/N}-1)}{N (e^{1/N}-1)}=1$\\
$M = 1$  $\blacksquare$\\
\end{spacing}
\end{center}
b) The expectancy is at its maximum when the probability of a single event is 1, then every other event probability is zero,  $P[\textbf{X}=x_j] = 1$ and $P[\textbf{X}=x_i,\forall i\neq j] = 0$\\
\begin{center}
\begin{spacing}{2.5}
Consider $S_N(\textbf{A})$ with $N \to \infty$, then for Theorem 1 $\beta_N = 1$\\
Then $\beta_N \cdot \sum \limits_{i=0}^{N-1}(e^{\alpha_i}-1) = 1 \cdot (0+...+0+(e-1))$\\
$M = e-1$  $\blacksquare$\\
\end{spacing}
\end{center}

\begin{proposition}

The syntropy $S_N(\textbf{A})$ is defined as the average expectancy value for any random variable $\textbf{A}$.
\end{proposition}

Consider that $A=\{ \alpha_0,\alpha_1,...,\alpha_n\}$ is a probability distribution and its expectancy is defined by:

$$ E(\textbf{A}) = \beta_N \{ (e^{\alpha_0}-1)+(e^{\alpha_1}-1)+...+(e^{\alpha_n}-1)\}$$

If $M$ trials of the variable $\textbf{A}$ are performed, then it is known that $\alpha_i = \frac{m_i}{M}$, where $m_i$ is the frequency with which each event $i$ appears. Then the average expectancy will be:

\begin{center}
\begin{spacing}{2.5}
$S_N(\textbf{A}) \triangleq \frac{E_M(\textbf{A})}{M}=\frac{\beta_N}{M}\{ m_0(e^{\alpha_0}-1)+m_1(e^{\alpha_1}-1)+...+m_n(e^{\alpha_n}-1)  \} $\\
By distributive law:\\
$=\beta_N\{ \frac{m_0}{M}(e^{\alpha_0}-1)+\frac{m_1}{M}(e^{\alpha_1}-1)+...+\frac{m_n}{M}(e^{\alpha_N}-1)  \}$\\
thus: \\
$S_N(\textbf{A})=\beta_N\{ \alpha_0(e^{\alpha_0}-1)+\alpha_1(e^{\alpha_1}-1)+...+\alpha_n(e^{\alpha_n}-1)  \}$\\
Resulting in:\\
 $S_N(\textbf{A}) \triangleq \beta_N \sum \limits_{i=0}^{N-1}\alpha_i(e^{\alpha_i}-1)$ $\blacksquare$\\
\end{spacing}
\end{center}

\end{multicols}

\clearpage

\subsection{Properties}

\subsubsection{Submultiplicativity and Supramultiplicativity}

To demonstrate that $S_N(\textbf{A})$ is submultiplicative or supramultiplicative as a function, we need to show that for any two probability values $p$ and $q$, the following inequality holds:

$$S_N(pq) \geq S_N(p)S_N(q) \quad \text{or} \quad S_N(pq) \leq S_N(p)S_N(q)$$

First lets consider:

$$S_N(pq) \geq S_N(p)S_N(q)$$

We know that:

$$S_N(p) = \frac{e^p - 1}{N(e^{1/N} - 1)} \quad \text{and} \quad S_N(q) = \frac{e^q - 1}{N(e^{1/N} - 1)}$$

Substituting these values into the inequality, we get:

$$\frac{e^{pq} - 1}{N(e^{1/N} - 1)} \geq \frac{e^p - 1}{N(e^{1/N} - 1)} \cdot \frac{e^q - 1}{N(e^{1/N} - 1)}$$

Applying Theorem 1, we get:

$$e^{pq} - 1 \geq (e^p - 1)(e^q - 1)$$

\begin{figure}[H]
\centering
\includegraphics[width=0.6\linewidth]{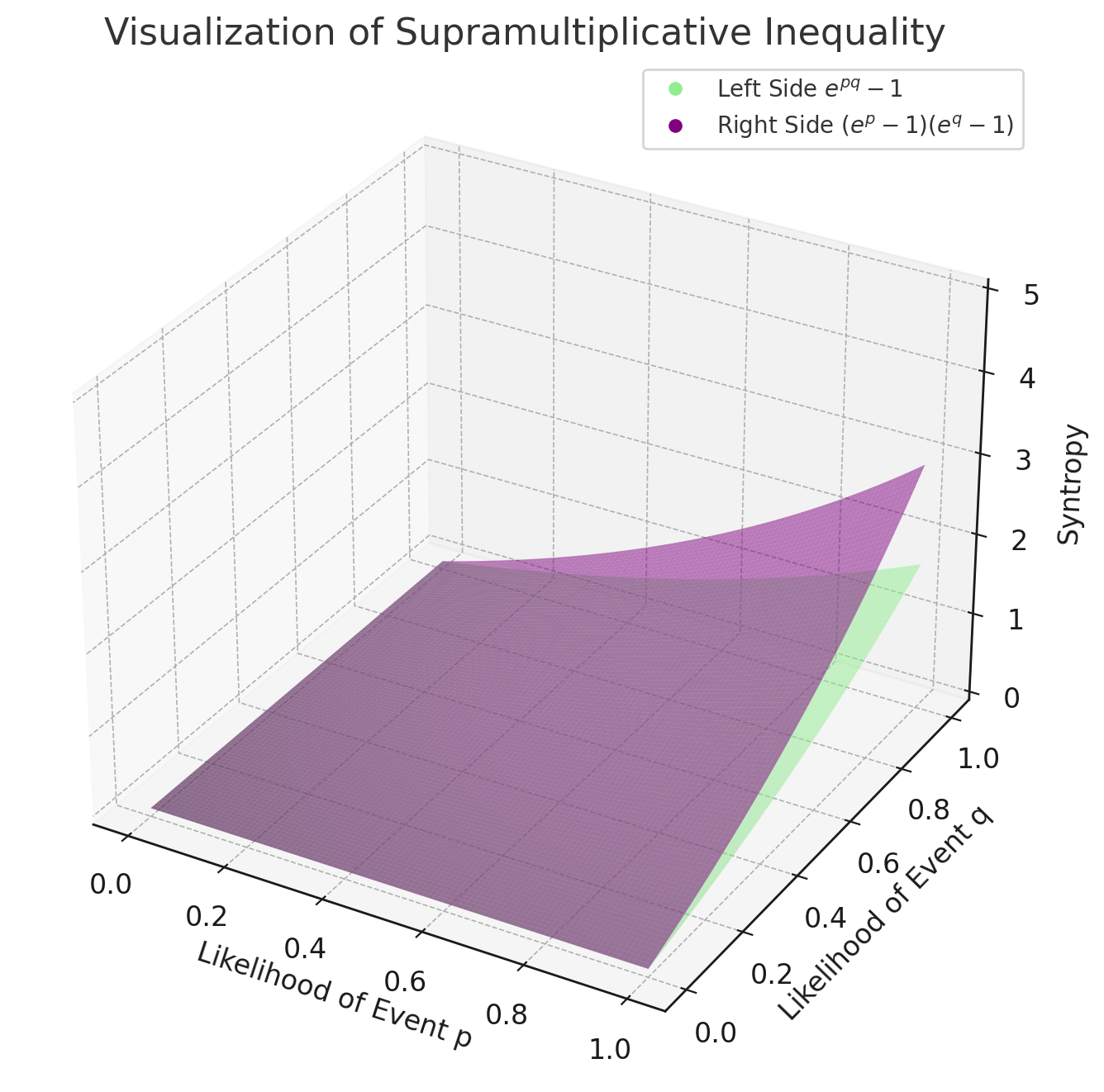}
\caption{This 3D visualization illustrates the supramultiplicative inequality comparison for the syntropy function \( S_N(\textbf{A}) \) within the domain of probabilities. The X-axis represents the likelihood of event p, the Y-axis the likelihood of event q, and the Z-axis the magnitude of syntropy. The light green surface corresponds to the left side of the inequality \( e^{pq} - 1 \), while the purple surface represents the right side \( (e^p - 1)(e^q - 1) \). The overlapping of the purple surface over the light green highlights the regions where the supramultiplicativity holds true.}
\label{SMS}
\end{figure}

This inequality proves that $S_N(\textbf{A})$ is both supramultiplicative and submultiplicative, but within $\mathbbm{R}[0,1)$ it is only supramultiplicative as shown in figure \ref{SMS}.

This inequality holds for all $p,q \in \mathbbm{R}[0,1)$, which proves that $S_N(\textbf{A})$ is supramultiplicative for any random variable.

\subsubsection{Subadditivity and Supradditivity}

To demonstrate that $S_N(\textbf{A})$ is subadditive or supraadditive as a function, we need to show that either:

1. $S_N(p+q) \leq S_N(p) + S_N(q)$ for all $p, q \in \textbf{A}$, which would mean that $S_N(\textbf{A})$ is subadditive.

2. $S_N(p+q) \geq S_N(p) + S_N(q)$ for all $p, q \in \textbf{A}$, which would mean that $S_N(\textbf{A})$ is supraadditive.

To demonstrate if its subadditive, we need to show that:

$$S_N(p+q) \leq S_N(p) + S_N(q)$$

Let $p$ and $q$ be real numbers, then:

$$S_N(p+q) = \frac{e^{p+q} - 1}{N(e^{\frac{1}{N}} - 1)}, \quad S_N(p) = \frac{e^{p} - 1}{N(e^{\frac{1}{N}} - 1)}, \quad S_N(q) = \frac{e^{q} - 1}{N(e^{\frac{1}{N}} - 1)}$$

Substituting the above expressions, we get:

$$\frac{e^{p+q} - 1}{N(e^{\frac{1}{N}} - 1)} \leq \frac{e^{p} - 1}{N(e^{\frac{1}{N}} - 1)} + \frac{e^{q} - 1}{N(e^{\frac{1}{N}} - 1)}$$

Multiplying both sides by $N(e^{\frac{1}{N}} - 1)$, we get:

$$e^{p+q} - 1 \leq e^{p} - 1 + e^{q} - 1$$

Simplifying:

$$e^{p+q} \leq e^{p}+e^{q}-1$$

Since $0 \leq p+q \leq 2$ and $e^t \geq 1 \forall t \in \mathbbm{R}(0,1)$, then:

$$e^2 \leq 2 e^1 -1$$

This is always false for all values of $x,y$, so we have shown that $S_N(x)$ is not subadditive.\\

For supraadditive we have:

$$e^{p+q} \geq e^{p}+e^{q}-1$$

\begin{figure}[H]
\centering
\includegraphics[width=0.6\linewidth]{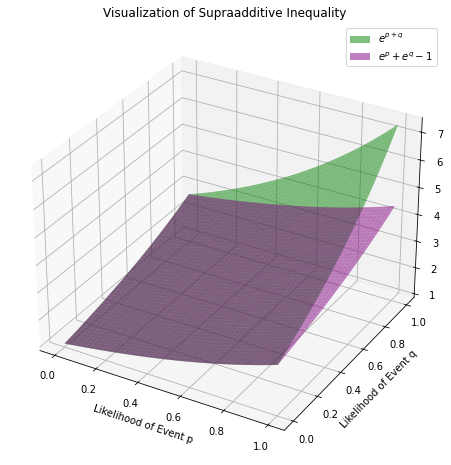}
\caption{This 3D visualization graphically represents the supraadditive inequality \(e^{p+q} \geq e^{p} + e^{q} - 1\). The X and Y axes represent the likelihoods of events p and q, respectively, within the domain \([0, 1]\). The green surface illustrates the left side of the inequality, \(e^{p+q}\), while the purple surface depicts the right side, \(e^{p} + e^{q} - 1\). The overlay of these surfaces visually demonstrates the regions where the inequality holds, emphasizing the supraadditive nature of this exponential function in the context of probabilistic events.}
\label{SAS}
\end{figure}

Which is true for all values of $p,q \in \mathbbm{R}(0,1)$, thus, demonstrating that $S_N(\textbf{A})$ is supraadditive as shown in figure \ref{SAS}.

\subsubsection{Complementary symmetry}

Complementary symmetry is a property in which the values of a function $f(x)$ at a point x and its complement $(1-x)$ are related in a symmetric way. In the case of $S_N(x)$, the complementary symmetry property states that $S_N(x) = S_N(1-x)$ for any probability mass function (pmf) $X$.

\begin{figure}[ht]
\centering
\includegraphics[width=0.8\linewidth]{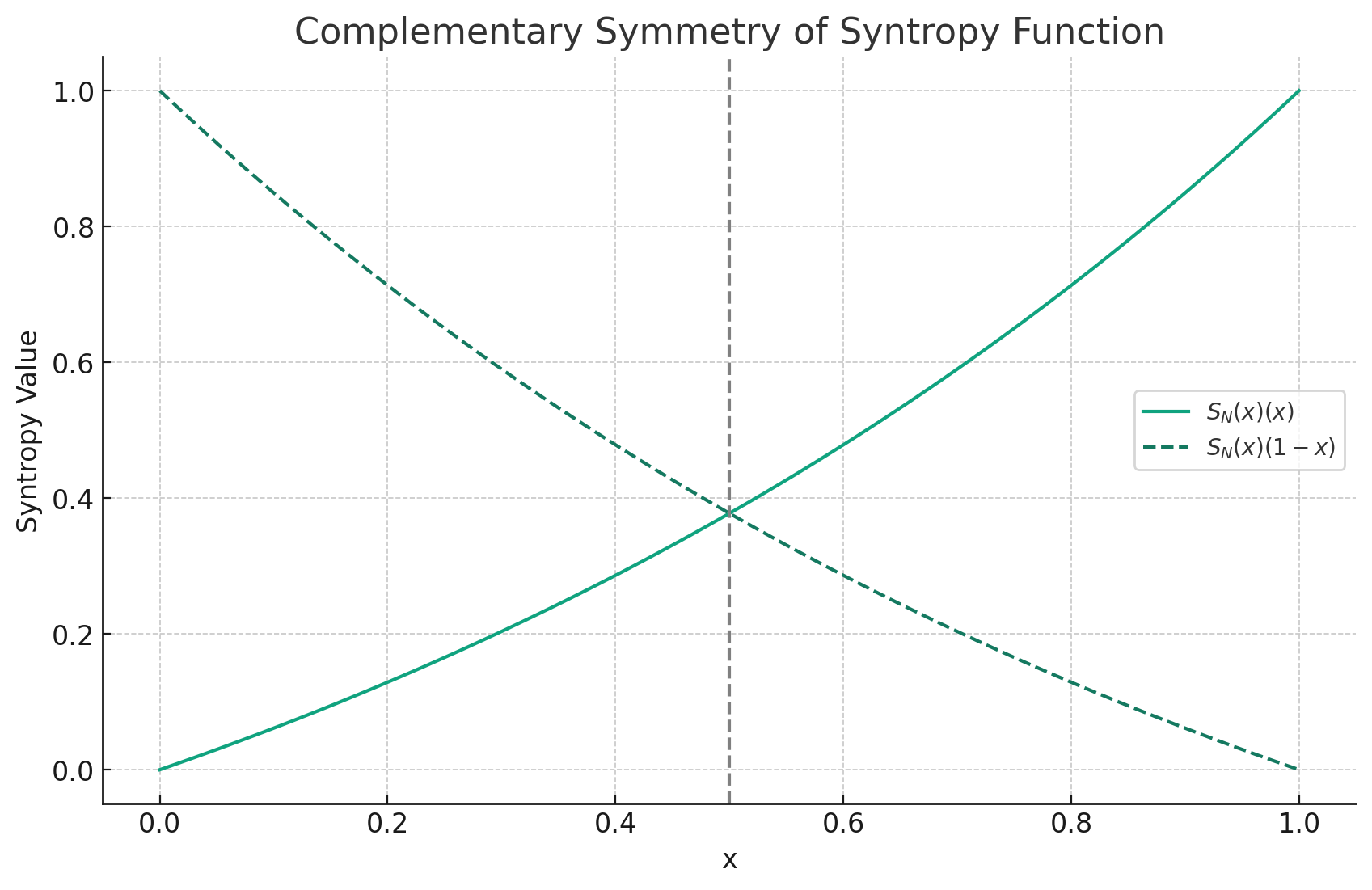}
\caption{Complementary Symmetry of the Syntropy Function \( S_N(x) \). The plot illustrates \( S_N(x) \) as a solid line and \( S_N(1-x) \) as a dashed line. The vertical gray dashed line at \( x = 0.5 \) serves as a reference to demonstrate the symmetry. The overlap and mirroring of the two curves around \( x = 0.5 \) highlight the complementary symmetry property, indicating that the values of \( S_N(x) \) at \( x \) and \( 1-x \) are equal for any probability mass function (pmf) \( X \).}
\label{symmetry}
\end{figure}

To illustrate this property, we can graph $S_N(x)$ against $S_N(1-x)$ for a given pmf. The resulting plot would be symmetric with respect to the line $x=0.5$, indicating that the values of $S_N(x)$ at $x$ and $1-x$ are equal as shown in Figure \ref{symmetry}.

Having complementary symmetry endows the following advantages to $S_N(x)$: It provides a way to check the validity of the calculation of $S_N(x)$, since the curve should be symmetric around the line $x=0.5$. Any deviation from this symmetry could indicate a calculation error or a problem with the underlying data; the complementary symmetry property allows for a more efficient computation of $S_N(x)$ in certain cases. Specifically, if a random variable has a symmetric probability distribution around $x=0.5$, then the values of $S_N(x)$ at $x$ and $1-x$ will be equal, and only half of the function needs to be computed; complementary symmetry allows for a more intuitive interpretation of $S_N(x)$, as it relates to the balance between order and disorder in a system. Specifically, if the curve is symmetric around $x=0.5$, it indicates that the system has a balance between order and disorder, whereas if it is skewed towards one side, it indicates a bias towards either order or disorder.

\subsubsection{Stationarity}

To show that \( S_N(x) \) is stationary, we need to prove that its value remains constant over time for a given probability distribution. Let's consider a probability distribution function \( f(x) \) and calculate \( S_N(x) \) for two different time intervals, \( t_1 \) and \( t_2 \), where \( t_2 > t_1 \).

For \( t_1 \), we have:
\[
S_N(f(x); t_1) = \frac{e^{f(x)} - 1}{N(e^{1/N} - 1)}
\]

Now, let's consider \( t_2 \). Since we are assuming that the probability distribution function \( f(x) \) is constant over time, it remains the same for \( t_2 \). Therefore, we have:
\[
S_N(f(x); t_2) = \frac{e^{f(x)} - 1}{N(e^{1/N} - 1)}
\]

Comparing both equations, we can see that they are identical, which means that \( S_N(x) \) is stationary with respect to time for a given probability distribution.

This property is useful in many applications, as it allows us to analyze the statistical properties of a process over time without having to track its evolution. We can simply calculate \( S_N(x) \) at different points in time and compare the results.

\begin{figure}[ht]
\centering
\includegraphics[width=0.8\linewidth]{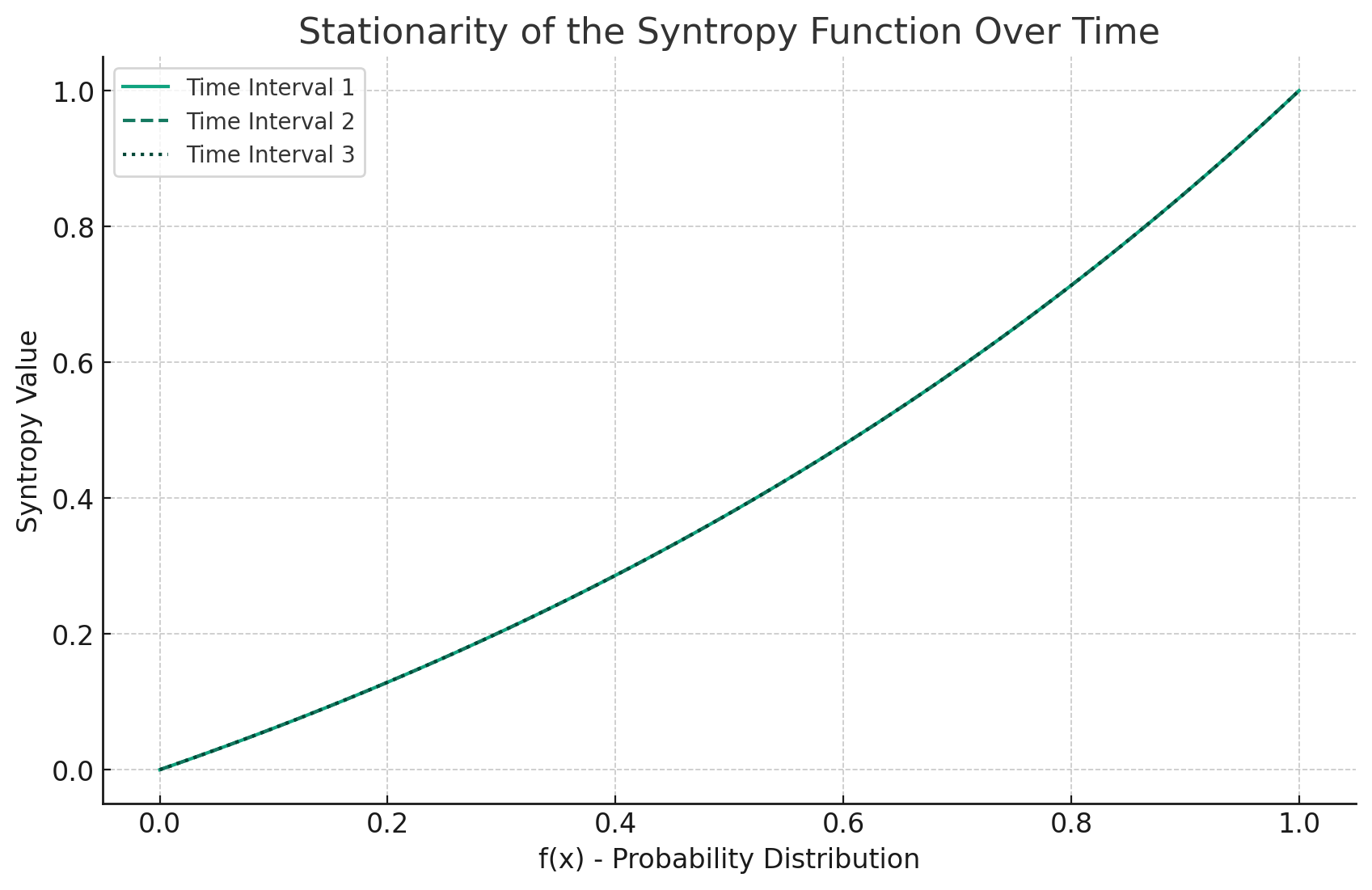}
\caption{Illustration of the Stationarity of the Syntropy Function \( S_N(x) \). The graph displays \( S_N(x) \) calculated for a consistent probability distribution function \( f(x) \) across three distinct time intervals, represented by different line styles: solid, dashed, and dotted. The congruence of these lines, irrespective of the time intervals, visually demonstrates the stationary nature of \( S_N(x) \). This consistency highlights that the syntropy value remains invariant over time as long as the underlying probability distribution \( f(x) \) does not change, emphasizing the temporal stability and dependability of \( S_N(x) \) in statistical analysis.}
\label{station}
\end{figure}

Figure \ref{station} effectively illustrates the stationarity of the syntropy function \( S_N(x) \) in the context of a constant probability distribution function \( f(x) \). In this visualization, we have depicted \( S_N(x) \) calculated for \( f(x) \) over three different time intervals, each marked by a distinct line style: solid, dashed, and dotted. Despite varying the time intervals, the three representations of \( S_N(x) \) are identical, clearly demonstrating that the syntropy function is stationary with respect to time. This means that as long as the underlying probability distribution \( f(x) \) remains constant, the value of \( S_N(x) \) does not change, regardless of the time at which it is evaluated. This characteristic of stationarity underscores the reliability and consistency of \( S_N(x) \) as an analytical tool in applications where the statistical properties of a process remain unchanged over time.

\subsubsection{Differentiability}

The syntropy function \( S_N(\alpha) \) is differentiable with respect to \( \alpha \) for all values of \( N \) except \( N=1 \), where it is not differentiable at \( \alpha=0 \). This can be seen from the formula:

\[
S_N(\textbf{A}) = \sum\limits_{i =1}^{N}\frac{e^{\alpha_i} - 1}{N(e^{1/N} - 1)}
\]

Taking the derivative with respect to \( x \) gives:

\[
\frac{dS_N(\textbf{A})}{d\textbf{A}} = \sum\limits_{i =1}^{N}\frac{e^{\alpha_i}}{N(e^{1/N} - 1)}
\]

This derivative is well-defined and continuous for all values of \( N \) except \( N=1 \), where the denominator becomes 0 at \( \alpha_i=0 \), making the derivative undefined.

Additionally, taking the derivative of \( S_\textbf{A}(N) \) with respect to \( N \) yields:

\[
\frac{dS_{\textbf{A}}(N)}{dN} =  \frac{e^{1/N} - 1 - \frac{1}{N}e^{1/N}}{N^3(e^{1/N} - 1)^2}\sum\limits_{i =1}^{N} (e^{\alpha_i}-1)
\]

This derivative shows how the syntropy function changes with respect to variations in \( N \). Similar to the derivative with respect to \( x \), it also encounters difficulties at \( N=1 \) and \( x=0 \) due to the denominator.

For \( N=1 \), as \( \alpha_i \) approaches 0, the function \( S_N(\textbf{A}) \) becomes increasingly steep, leading to a discontinuity in both its derivatives with respect to \( \alpha_i \) and \( N \). Therefore, while \( S_N(\textbf{A}) \) exhibits differentiability across a wide range of \( N \) and \( \alpha_i \), it encounters critical points at \( N=1 \) and \( \alpha_i=0 \), where the derivatives do not exist.

\subsubsection{Sensitivity to intial conditions}

$S_N(\textbf{A})$ is not sensitive to initial conditions. This can be seen from the formula, which does not involve any derivatives or integrals, and depends only on the value of $x$ and the constant $N$. Therefore, small changes in the initial value of $x$ will not result in large changes in the value of $S_N(\textbf{A})$.

\subsubsection{Robustness}

$S_N(\textbf{A})$ is robust to noise and outliers. This is because it is based on the spectral properties of the signal, which are relatively insensitive to noise and outliers. In addition, the normalization factor $\beta_N$ ensures that the measure is scaled appropriately, regardless of the magnitude of the signal. However, the measure is sensitive to the choice of the spectral window used in the calculation, and different window functions may give different results. Therefore, it is important to choose an appropriate window function based on the characteristics of the signal being analyzed.

\subsubsection{Computational efficiency}

The computational efficiency of $S_N(\textbf{A})$ is relatively good compared to some other complexity measures, such as Kolmogorov Complexity or Fractal Dimension. This is because the computation of $S_N(\textbf{A})$ involves only simple arithmetic operations, such as exponentiation, subtraction, and division, which can be efficiently implemented in most programming languages and hardware architectures. The big O notation for $S_N(\textbf{A})$ is $O(e^{\alpha_i}/\beta_N)$.

\subsubsection{Generalization to higher dimensions}

The $S_N(\textbf{A})$ function can be extended to higher dimensions by considering joint probability distributions instead of individual probabilities. For example, for a two-dimensional system with joint probability distribution $p(\alpha,\gamma)$, the spectral syntropy can be defined as $S_N(\textbf{A})\triangleq S(N;\alpha,\gamma) = \frac{e^{\alpha+\gamma}-1}{N(e^{2/N}-1)}$. This measure provides a quantification of the joint complexity and integration of the system.

However, the generalization of $S_N(\textbf{A})$ to higher dimensions faces several challenges. One challenge is the curse of dimensionality, which refers to the exponential increase in the size of the probability space as the number of dimensions increases. This can make the computation of $S_N(\textbf{A})$ intractable for high-dimensional systems.

Another challenge is the interpretation of the results, as the higher-dimensional spectral syntropy measures may not have a clear physical or biological meaning. Therefore, the use of higher-dimensional spectral syntropy measures requires careful consideration of the specific application and the interpretation of the results.

\subsection{Interaction with Shannon's Entropy}

\subsubsection{Equilibrium Point}
\begin{figure}[ht]
\centering
\includegraphics[width=0.75\textwidth]{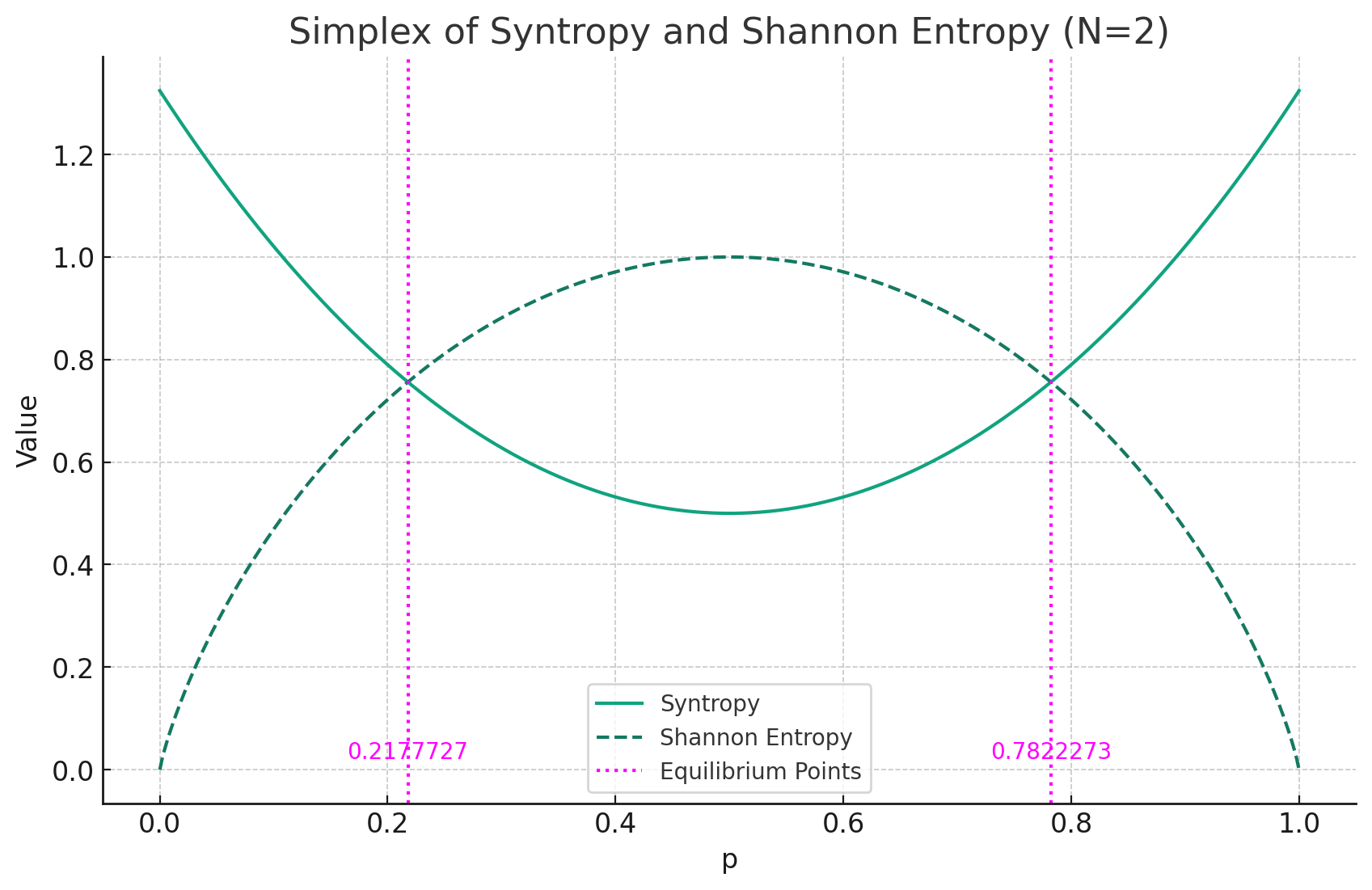}
\caption{The simplex of \( N=2 \) showing the equilibrium point where the measures of Shannon entropy \( H(p) \) and syntropy \( S(p) \) intersect.}
\label{simplexE}
\end{figure}

The equilibrium point in the context of entropy and syntropy of a probability distribution represents a state of balance between the unpredictability and the inherent order within a system. For a simplex of \( N = 2 \), this balance point is graphically represented by the intersection of the Shannon entropy function \( H(p) \) and the syntropy function \( S(p) \), as depicted in the Figure \ref{simplexE}.

The Shannon entropy \( H(p) \) for a probability \( p \) and its complement \( 1-p \) is defined as:

\[ H(p) = -p \log_N(p) - (1-p) \log_N(1-p) \]

The syntropy \( S(p) \) is similarly defined as:

\[ S(p) = p\frac{e^p - 1}{N(e^{1/N} - 1)} + (1-p)\frac{e^{1-p} - 1}{N(e^{1/N} - 1)} \]

The equilibrium point \( p \) is found where \( H(p) = S(p) \). Mathematically, this is expressed as finding the root of the equation \( S(p) - H(p) = 0 \). The process to determine this involves numerical analysis, typically employing a root-finding algorithm such as the bisection method. The bisection method iteratively narrows down the interval within which the equilibrium point lies, ensuring that the function changes sign over the interval, until the width of the interval is less than a specified tolerance level, here set to \( 1 \times 10^{-16} \).

The Java code used for this numerical procedure is outlined in the annex to this article. It implements the syntropy and Shannon entropy functions, along with the bisection method to find the equilibrium point.

\begin{figure}[ht]
\centering
\includegraphics[width=0.75\textwidth]{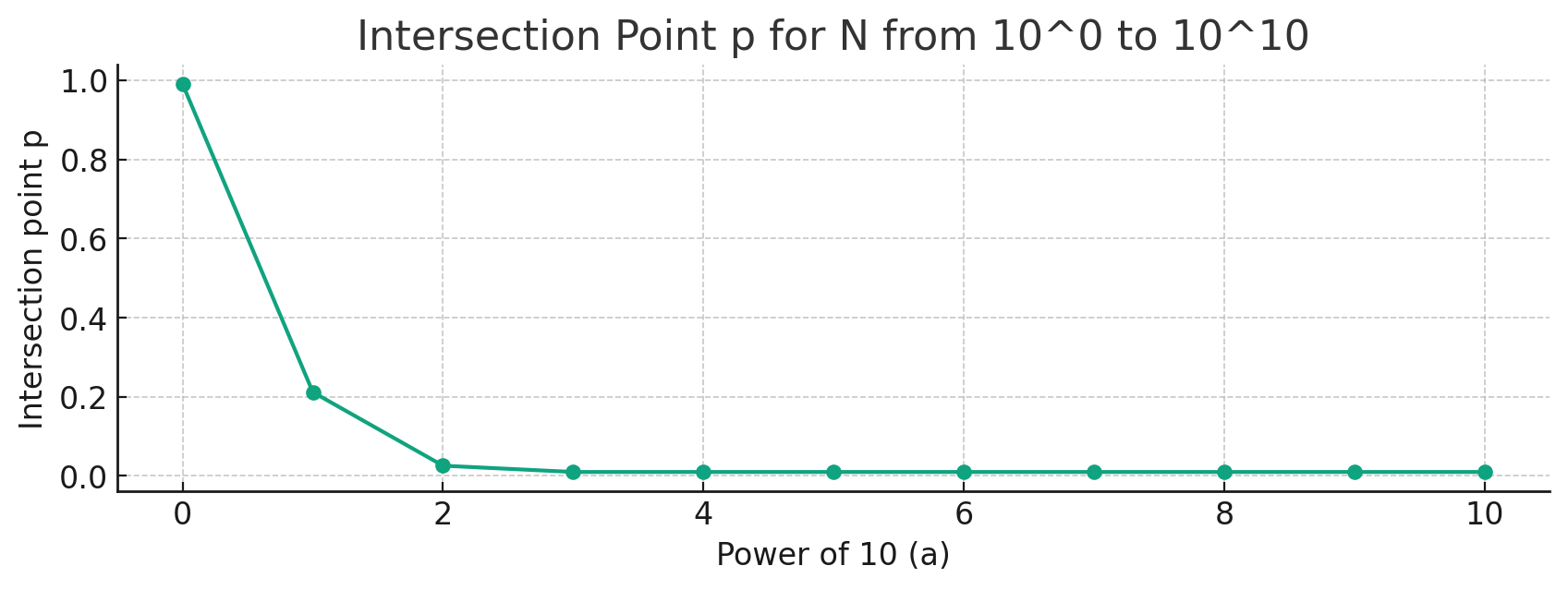}
\includegraphics[width=0.75\textwidth]{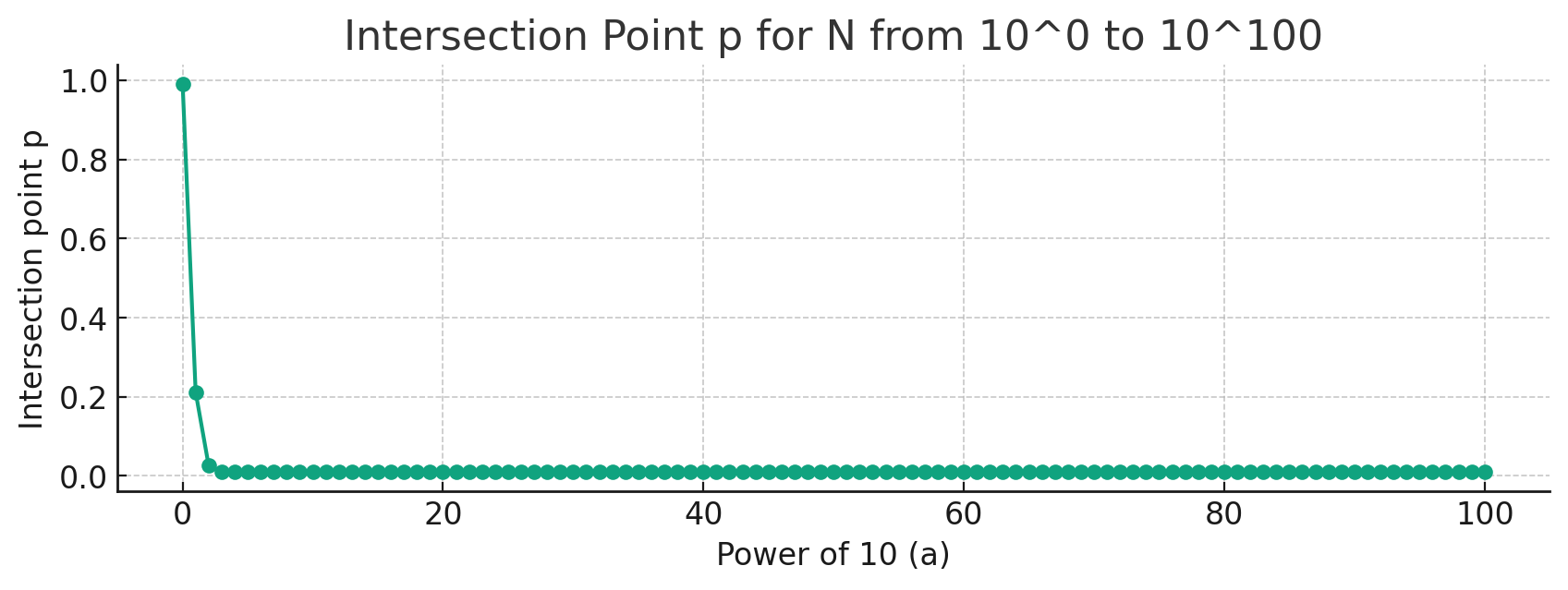}
\caption{The top graph illustrates the relationship between the intersection point \( p \) and the system size \( N \) (ranging from \( 10^0 \) to \( 10^{10} \)). The bottom graph extends this analysis, presenting the shift in \( p \) for \( N \) from \( 10^0 \) to \( 10^{100} \). Both plots reveal the dynamic balance between entropy and syntropy in systems of varying complexity, highlighting how the intersection point \( p \) decreases as \( N \) increases exponentially.}
\label{syntropy_entropy_intersection}
\end{figure}

Preliminary findings suggest an apparent convergence of the equilibrium point \( p = 0.009999999776482638\) as \( N \) grows large as shown in Figure \ref{syntropy_entropy_intersection}. The significance of this convergence indicates that as the system size \( N \) increases, the balance between entropy and syntropy converges to a specific probability \( p \), potentially signifying a universal characteristic of complex systems' behavior at scale.

\subsection{Distance to Equilibrium}
The equilibrium between entropy and syntropy is characterized by a unique point where the system's unpredictability and inherent order are in perfect balance. To quantify the deviation from this state, we introduce the concept of the Distance to Equilibrium, denoted as \( DAE(x) \). This measure is based on the normalized values of entropy \( H(x) \) and syntropy \( S(x) \), providing a standardized metric for comparing different systems or states within a system.

The \( DAE(x) \) is defined as the difference between the normalized syntropy and a scaled normalized entropy, where the scaling factor is \( \frac{1}{e-1} \), accounting for the exponential nature of syntropy relative to the logarithmic nature of entropy. The normalization process involves subtracting the minimum possible value of the function from the actual value and then dividing by the range of the function. This procedure ensures that both \( H(x) \) and \( S(x) \) are scaled to a [0,1] interval, facilitating a direct comparison. Mathematically, \( DAE(x) \) is given by:

\begin{equation*}
  DAE(x) = \text{Norm}(S(x)) - \frac{1}{e-1} \times \text{Norm}(H(x))
\end{equation*}

where the normalized syntropy \( \text{Norm}(S(x)) \) is defined as:

\begin{equation*}
  \text{Norm}(S(x)) = \frac{S(x) - \text{min}(S(x))}{\text{max}(S(x)) - \text{min}(S(x))}
\end{equation*}

and the normalized entropy \( \text{Norm}(H(x)) \) is similarly defined:

\begin{equation*}
  \text{Norm}(H(x)) = \frac{H(x) - \text{min}(H(x))}{\text{max}(H(x)) - \text{min}(H(x))}
\end{equation*}

By constructing \( DAE(x) \) in this manner, we can observe how a system's current state relates to its equilibrium. A \( DAE(x) \) value close to zero indicates proximity to equilibrium, while larger absolute values signal a greater distance from this point of balance. It is through this lens that we can assess the dynamical nature of complex systems and their tendency to either maintain homeostasis or evolve towards new states of order or chaos.

\section{Time Series Application}

\subsection{Syntropy}

Shannon's Entropy, a fundamental concept in information theory, is adeptly applied to quantify the uncertainty or randomness within time series data. This measure is instrumental in various domains, including but not limited to, telecommunications, finance, and environmental science, providing insights into the underlying dynamics and complexity of temporal data sequences. Nevertheless, measuring certainty and order is not a lineal or complementary task, and so, syntropy complements that part

\subsubsection{Definition}

Given a discrete time series \(X = \{x_1, x_2, \ldots, x_N\}\), where each \(x_i\) represents a value at the \(i\)-th time point and \(N\) is the total number of observations, then time dependent syntropy (\textbf{TDS}) can be defined as:

\begin{equation}
   S_N(X) = \beta_N \sum_{i=1}^{n} p(x_i) (e^{ p(x_i)}-1)
\end{equation}

where \(p(x_i)\) denotes the probability of occurrence of the \(i\)-th value in the time series, \(n\) is the number of unique values and $\beta_N$ is the normalizing factor.

\subsubsection{Components}

\paragraph{Probability Distribution Estimation}

For a discrete time series \(X = \{x_1, x_2, \ldots, x_N\}\), the first step in computing \textbf{TDS} is to estimate the probability distribution of the observed values. This involves counting the frequency of each unique value \(x_i\) in the series and then normalizing these frequencies to obtain probabilities:

\begin{equation}
    p(x_i) = \frac{\text{Frequency of } x_i}{N},
\end{equation}

where \(N\) is the total number of observations in the time series. This process converts the raw time series data into a probability distribution, where \(p(x_i)\) represents the likelihood of encountering the value \(x_i\) in the dataset.

\paragraph{Information Content}

The information content of each unique value \(x_i\) in the time series, given its probability \(p(x_i)\), is quantified by:

\begin{equation}
    I(x_i) = \beta_N  (e^{ p(x_i)}-1)
\end{equation}

Contrary to Shanon's Entropy, this measure reflects the amount of "expectancy" or "certainty" associated with observing the value \(x_i\), with more probable events contributing more to the overall syntropy.

\subsubsection{Calculation}

The \textbf{TDS} for the time series is then computed by summing the weighted information content across all unique values:

\begin{equation}
    S(X) = \sum_{i=1}^{n} p(x_i) I(x_i) = \beta_N \sum_{i=1}^{n} p(x_i) (e^{ p(x_i)}-1),
\end{equation}

where \(n\) is the number of unique values in the discretized time series or it can be a set of proposed threshold values.

\subsubsection{Interpretation}

The \textbf{TDS} serves as a measure of the predictability or coherence inherent in a time series. A lower syntropy indicates a more complex and less predictable series, characterized by a more uniform distribution of values. Conversely, higher syntropy suggests more predictability and less randomness, often implying repetitive or regular patterns within the data.

\subsubsection{Graphical Visualization}

\begin{figure}[ht]
    \centering
    \includegraphics[width=0.8\textwidth]{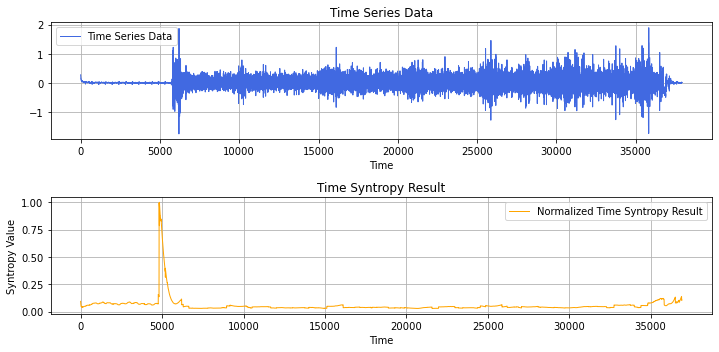}
    \caption{Example visualization of Syntropy applied to a sEMG registry, highlighting periods of varying complexity and predictability.}
    \label{syntropy_visualization}
\end{figure}

Visualizing \textbf{TDS} in the context of a time series might involve plotting the distribution of the unique values or changes in entropy over time, particularly if sliding windows are used to analyze dynamic changes in complexity. In figure \ref{syntropy_visualization} display the time series with annotations indicating periods of high versus low syntropy, providing a visual representation of the data's changing complexity and predictability.

\section{Spectral Application}

Spectral Syntropy (\textbf{SS}) is a measure derived from \textbf{TDS}, applied to the power spectrum of a time series. It quantifies the distribution and uniformity of power across various frequency components, offering insights into the regularity and complexity of the signal. This metric will be particularly valuable in signal processing, biomedical signal analysis, and other fields where understanding the frequency domain characteristics of a time series is crucial.

\subsection{Definition}

For a given time series \(X = \{x_1, x_2, \ldots, x_N\}\), the \textbf{SS} is defined based on the normalized power spectral density (\textbf{PSD}) of the series. Let \(P(f)\) denote the normalized \textbf{PSD} of \(X\), computed over a set of frequencies \(f\). The Spectral Entropy is then given by:

\begin{equation}
    S_x = \beta_N \sum_{i=1}^{n} p(\omega_i) (e^{ p(\omega)}-1),
\end{equation}

where \(p(\omega_i)\) represents the normalized power at frequency \(\omega_i\), ensuring that \(\sum_{i=1}^{n} p(\omega_i) = 1\), and \(n\) is the number of discrete frequencies in $P(f)$.

\subsection{Components}

\paragraph{Power Spectral Density}

The Power Spectral Density (\textbf{PSD}) of a time series \(X = \{x_1, x_2, \ldots, x_N\}\) is a measure of the signal's power content versus frequency. It is defined as:

\begin{equation}
    P(\omega) = \left|\mathcal{F}\{x(t)\}\right|^2,
\end{equation}

where \(\mathcal{F}\{x(t)\}\) denotes the Fourier Transform of the time series \(x(t)\), and \(|\cdot|^2\) represents the squared magnitude, ensuring that \(P(\omega_i)\) denotes the power of the magnitude of said frequency. The \textbf{PSD} is estimated over a discrete set of frequencies \(\omega_i\), typically obtained via the Fast Fourier Transform (FFT) or Discrete Fourier Transform (\textbf{DFT}) for computational efficiency.

\paragraph{Normalization of PSD}

To apply \textbf{SS} to the \textbf{PSD}, it must first be normalized to resemble a probability mass function, where the sum of all probabilities equals 1. This is achieved by dividing the power at each frequency by the total power across all frequencies, yielding:

\begin{equation}
    p(f_i) = \frac{P(\omega_i)}{\sum_{j=1}^{n} P(\omega_j)},
\end{equation}

where \(P(\omega_i)\) is the power at frequency \(\omega_i\), and \(n\) is the total number of discrete frequencies in the PSD. This normalization ensures that \(\sum_{i=1}^{n} p(\omega_i) = 1\), satisfying the requirement for \(p(\omega_i)\) to represent a probability distribution. When analyzing a large time series through multiple windows, to ensure better results let's assume that every $\omega_i$ its an event that will change through every window, so instead of normalizing against the window, one should normalize against the window with the most energy or the maximum energy of the channel that is being analyzed.
.
\subsection{Interpretation}

The \textbf{SS} provides a measure of the predictability and complexity of a time series from a frequency domain perspective. A lower \textbf{SS} value indicates a more concentrated or peaky \textbf{PSD}, suggesting a time series dominated by fewer frequency components and potentially more regular. Higher \textbf{SS} values indicate a more uniform distribution of power across frequencies, suggesting a more complex and less predictable signal.

\subsection{Graphical Visualization}

Visualizing \textbf{SS} typically involves plotting the \textbf{PSD} of the time series alongside the entropy value. This visualization can highlight the frequency components contributing to the overall entropy and provide insights into the signal's complexity as seen in Figure \ref{spectral_syntropy_visualization}.

\begin{figure}[h]
    \centering
    \includegraphics[width=0.8\textwidth]{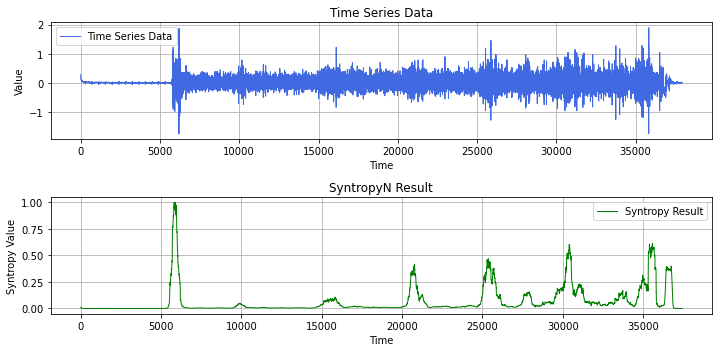}
    \caption{Illustration of Spectral Entropy computation, showing the PSD of a time series and the corresponding entropy value.}
    \label{spectral_syntropy_visualization}
\end{figure}
\section{Vectorial Transformation and Correlation Analysis using Spectral Syntropy}

This algorithm leverages the concept of \textbf{SS} to analyze the syntropy contribution across frequency components in a time series transformed through the Short-Time Fourier Transform (\textbf{STFT}) or Gabor Transform (\textbf{GT}). It provides insights into the regularity and complexity of the signal by applying a non-linear transformation to the probabilities associated with these frequency components, represented in a matrix form. This process involves normalization and correlation analysis to uncover linear relationships between the transformed components.

\subsection{Definition}

The algorithm is designed to analyze a time series by examining its syntropy components obtained from the \textbf{SS} or \textbf{TDS}. These components are then aggregated in a matrix $\{S_x\}_{mxn}$ . The process involves normalization of this matrix's rows and calculation of correlation coefficients between the rows to form a correlation matrix, which uncovers the linear relationships between the transformed elements.

\subsection{Components}

\paragraph{Frequency Components and Probabilities}

For frequency components derived from \textbf{STFT} or \textbf{GT}, the probability associated with a component $x_i$ is estimated based on its relative power within the spectrum:
\begin{equation}
    P(x_i) = \frac{|X(x_i)|^2}{\sum_{j=1}^{m} |X(x_j)|^2}
\end{equation}
where $|X(x_i)|^2$ represents the power of the frequency component $x_i$.

Alternatively, for directly observed values in the time series, the probability $p(x_i)$ of each unique value $x_i$ is determined by its frequency of occurrence:
\begin{equation}
    P(x_i) = \frac{\text{Frequency of } x_i}{a}
\end{equation}
where $a$ is the total number of observations in the time series, and "Frequency of $x_i$" is the count of occurrences of the value $x_i$.

\paragraph{Syntropy Transformation}

Initially, the concept of Spectral Syntropy is formulated as an aggregate measure, summing the contributions of all transformed probabilities associated with the frequency components or observed values within a data set:

\begin{equation}
    S_x = \sum_{i=0}^{m} I(P(x_i))
\end{equation}

where $I(P(x_i))$ represents the information content from the Syntropy transformation applied to each probability $P(x_i)$, and is defined by:

\begin{equation}
    I(P(x_i)) = \beta_N P(x_i)(e^{P(x_i)} - 1)
\end{equation}

In this aggregate form, $S_x$ captures the overall spectral diversity and complexity by combining the contributions from all components or values. However, for the purpose of constructing the matrix $\{S_x\}_{mxn}$ and conducting a detailed analysis of the relationships between different components or values, it is essential to isolate and utilize the individual transformed probabilities.

Hence, the transformation from the aggregate measure to individual elements is realized by considering each $I(P(x_i))$ independently, as opposed to summing them. This leads to the definition of each matrix element $S_{x,i}$ as:
\begin{equation}
S_{x,i} \triangleq I(P(x_i)) \quad \text{for  } i = 1, 2, \ldots, m
\end{equation}

This step emphasizes the individual contribution of each frequency component or observed value to the Spectral Syntropy, represented by $S_{x,i}$. The inclusion of the normalization constant $\beta_N$ within each transformation ensures that these contributions are appropriately scaled, allowing for a consistent and meaningful comparison across the matrix.

By isolating the transformed probabilities as individual elements $S_{x,i}$ in the matrix, the algorithm facilitates a granular analysis of the spectral characteristics, enabling the examination of linear relationships between different components or values through subsequent normalization and correlation analysis.

\paragraph{Matrix Representation}

The matrix $\{S_x\}_{mxn}$ is constructed by assigning Spectral Syntropy values to its elements, where each element $S_{i,n}$ is defined by the Spectral Syntropy transformation applied to the probability associated with a specific frequency component or observed value within the $n$-th window:
\begin{equation}
    S_{i,n} = I(P(x_i^n)) \quad \text{for } i = 1, 2, \ldots, m \text{ and } n = 1, 2, \ldots, N
\end{equation}

Here, $i$ indexes the frequency components or observed values, and $n$ serves as the window index, reflecting the sequential analysis of the dataset through sliding windows in either the \textbf{STFT}, \textbf{GT} or other spectral or time-domain analysis. Each row in the matrix corresponds to a transformed frequency component or observed value, and each column $n$ represents a distinct window of analysis, capturing the temporal evolution of the spectral characteristics.

This matrix $\{S_x\}_{mxn}$, with its elements reflecting the Spectral Syntropy within each window, enables a detailed examination of the changes and relationships in the spectral content over time or across different segments of the dataset. By analyzing this matrix, one can gain insights into the dynamics and variations in the spectral characteristics, facilitating a deeper understanding of the underlying processes or signals.

\paragraph{Normalization}

To ensure comparability across different windows and frequency components, each row of the matrix $\{S_x\}_{mxn}$ undergoes normalization. This process adjusts the values such that each row $R_i$, corresponding to a specific frequency component across all windows, has a mean of 0 and a standard deviation of 1. Mathematically, the normalized value $\hat{S}_{i,n}$ is given by:
\begin{equation}
    \hat{S}_{i,n} = \frac{S_{i,n} - \mu_{R_i}}{\sigma_{R_i}}
\end{equation}
where $\mu_{R_i}$ and $\sigma_{R_i}$ are the mean and standard deviation of the $i$-th row, respectively. This normalization step is essential for removing scale discrepancies and ensuring that the subsequent analysis is not biased by the magnitude of the spectral syntropy values. Figure \ref{spectral_syntropy_visualization} shows the resulting vectors as a spectrogram.

\begin{figure}[h]
    \centering
    \includegraphics[width=0.8\textwidth]{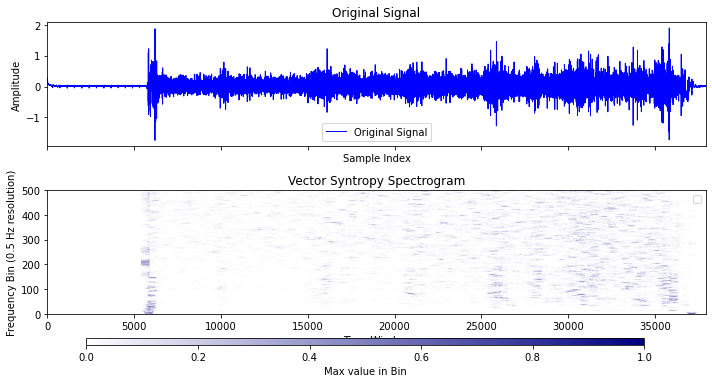}
    \caption{Visual representation of the correlation matrix and the effect of normalization on Spectral Syntropy values.}
    \label{spectral_syntropy_visualization}
\end{figure}

\paragraph{Correlation Analysis}

After normalization, correlation analysis is performed to explore the linear relationships between different frequency components or observed values across the windows. The Pearson correlation coefficient $\rho_{R_i, R_j}$ between any two normalized rows $\hat{R}_i$ and $\hat{R}_j$ is calculated as follows:
\begin{equation}
    \rho_{R_i, R_j} = \frac{\sum_{n=1}^{N} (\hat{S}_{i,n} - \bar{\hat{R}}_i)(\hat{S}_{j,n} - \bar{\hat{R}}_j)}{\sqrt{\sum_{n=1}^{N} (\hat{S}_{i,n} - \bar{\hat{R}}_i)^2} \sqrt{\sum_{n=1}^{N} (\hat{S}_{j,n} - \bar{\hat{R}}_j)^2}}
\end{equation}
where $\bar{\hat{R}}_i$ and $\bar{\hat{R}}_j$ are the means of the normalized rows $\hat{R}_i$ and $\hat{R}_j$, respectively. This correlation coefficient ranges from -1 to 1, where 1 indicates a perfect positive linear relationship, -1 indicates a perfect negative linear relationship, and 0 indicates no linear relationship. The resulting correlation matrix, composed of all pairwise correlation coefficients, provides a comprehensive view of the interdependencies among the spectral characteristics of the dataset over the analyzed windows.

\paragraph{Correlation Matrix Construction}

The correlation matrix, denoted as \(\mathbf{C}\), is constructed from the Pearson correlation coefficients obtained between every pair of normalized rows in the matrix \(\{S_x\}_{mxn}\). Each element \(C_{i,j}\) within the correlation matrix is defined as the correlation coefficient between the \(i\)-th and \(j\)-th frequency components or observed values across all windows, formalized as:
\begin{equation}
    C_{i,j} = \rho_{R_i, R_j} \quad \text{for } i, j = 1, 2, \ldots, m
\end{equation}

The diagonal elements of \(\mathbf{C}\), where \(i = j\), are always 1, reflecting the perfect correlation of each component with itself. The off-diagonal elements, where \(i \neq j\), range from -1 to 1, representing the degree and direction of linear correlation between different components.

\subsection{Interpretation}

The correlation matrix \(\mathbf{C}\) provides a comprehensive view of the linear relationships among the transformed frequency components or observed values. Elements in \(\mathbf{C}\) close to 1 or -1 indicate strong positive or negative linear correlations, respectively, suggesting a significant co-variation between the corresponding components across the analyzed windows. Values near 0 imply a weak or non-existent linear relationship.

This matrix is instrumental in identifying clusters of components that exhibit similar behavior over time, highlighting unique or divergent patterns, and offering insights into the underlying dynamics of the dataset. Such analyses are crucial for further investigations and model development aimed at understanding the processes influencing the spectral characteristics.

\subsection{Graphical Visualization}

\begin{figure}[h]
    \centering
    \includegraphics[width=0.8\textwidth]{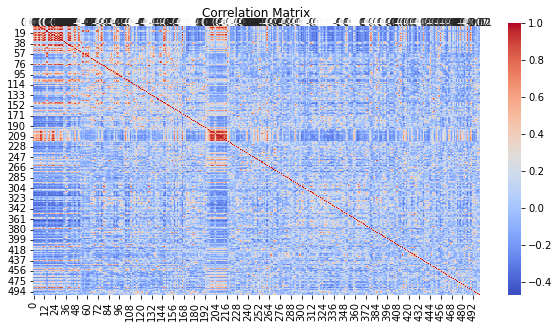}
    \caption{Visual representation of the correlation matrix and the effect of normalization on Spectral Syntropy values.}
    \label{correlation}
\end{figure}

Graphical visualization will include plotting the correlation matrix and the distributions of the transformed probabilities before and after normalization, offering a visual interpretation of the frequency domain characteristics and their relationships as shown in Figure \ref{correlation}.

\clearpage
\section{Dynamic Systems Application}

Dynamic systems exhibit complex interactions over time, where Syntropy provides insights into the constructive order within these systems.

\subsection{Dynamic Syntropy}

Given a state matrix $\textbf{X} = [x_{i,k}]$ for a dynamic system, with indices $i$ and $k$, Dynamic Syntropy for column $k$, denoted as $DynSyn(X_k)$, quantifies the ordered interactions.

\paragraph{Definition}

Dynamic Syntropy for a specific column $k$ of $\textbf{X}$ is defined as:

\begin{equation}
    DynSyn(X_k) = \sum_{i=0}^{\infty} \Phi_{i,k}(e^{\Phi_{i,k}} - 1), \quad \text{where } \Phi_{i,k} \neq 0
\end{equation}

This expression incorporates a measure $\Phi_{i,k}$ for each state in column $k$, reflecting the ordered interactions.

\paragraph{Measure $\Phi_{i,k}$}

For state $x_{i,k}$ in column $k$, $\Phi_{i,k}$ is given by:

\begin{equation}
    \Phi_{i,k} = \frac{\sum\limits_{j=0}^{\infty} \Theta(\alpha(x_{i,k},x_{j,k+1})>\beta(x_{i,k},x_{j,k-1}))}{C_k \cdot C_{k+1}}
\end{equation}

Here, $\Theta$ represents the Heaviside function, which equals 1 when its argument is positive and 0 otherwise.

\paragraph{State Counts $C_k$ and $C_{k+1}$}

The counts of non-zero states in columns $k$ and $k+1$, necessary for normalizing $\Phi_{i,k}$, are defined as:

\begin{equation}
    C_k = \sum_{j=0}^{\infty} \Theta(x_{j,k})
\end{equation}

\paragraph{Distance Functions}
The distance functions $\alpha$ and $\beta$ are defined as follows:\\

The $\alpha$ function measures the absolute difference between state $x_{i,k}$ and each state in column $k+1$.
\begin{equation}
 \alpha(x_{i,k},x_{j,k+1})=|x_{i,k} - x_{j,k+1}| \quad \text{for} \quad j = \{ 1,2,3, ... \infty\}
\end{equation}

The $\beta$ function finds the minimum difference between state $x_{i,k}$ and states in column $k-1$, excluding zero states, and $r$ is a tolerance factor to account for the natural growth of the system (In the example provided $r = 1.5$).
\begin{equation}
 \beta(x_{i,k},x_{j,k-1})= r* \min_{j=0,x_{j,k-1} \neq 0,x_{i,k}\neq 0}|x_{i,k}-x_{j,k-1}| \quad \text{for}\quad  j = \{ 1,2,3, ... \infty\}
\end{equation}

\subsection{Interpretation}

Dynamic Syntropy, through $DynSyn(X_k)$, captures the orderly and constructive patterns within a dynamic system, indicating the degree of coherence and potential predictability of its dynamics.

\subsection{Graphical Visualization}

\begin{figure}[h]
     \centering
     \includegraphics[width=0.8\textwidth]{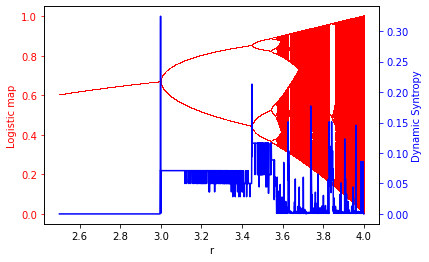}
     \caption{Visual representation of Dynamic Syntropy, illustrating the ordered interactions within the system.}
     \label{dynamic_syntropy_visualization}
 \end{figure}

Visualizing Dynamic Syntropy can highlight the structured interactions between states, offering a deeper understanding of the system's syntropic characteristics as shown in Figure \ref{dynamic_syntropy_visualization}.

\section{Conclusion}

In this work, we introduced theoretical construct of "expectancy," countering statistical surprise. Syntropy, derived from this expectancy, serves as an antithesis to Shannon's entropy, with no current ties to physical phenomena. While we've highlighted critical theorems and properties, ongoing research will expand these foundations. Future computational advancements will further scrutinize the equilibrium point and its implications for chaos and order, hinting at deeper interplays within complex systems. Additionally, the conceptual 'distance to equilibrium' shows promise in biological contexts, meriting further investigation. As syntropy complements entropy, exploring its variants remains a vital, rigorous endeavor, with vectorial transformations unveiling potential hidden intricacies within data signals.

\section{Discussion}

In this article, we introduced a function$phi$, chosen for its distinctive monotonic and injective properties, to serve as a theoretical measure of expectancy, in contrast to statistical surprise. This foundational concept underpins our exploration of syntropy, a measure designed to counterbalance the principles of Shannon's entropy and, in the same way as Shannon's Entropy, it does not directly correlate with any known physical phenomena.\\

The development of theorems and propositions for syntropy remains an active area of inquiry, with the current focus on those pivotal for the applications and propositions presented herein. As the utility of syntropy expands, we anticipate the emergence of additional properties that will necessitate further investigation and elucidation.\\

Computational advancements are expected to enhance our understanding of the equilibrium point explored in this study, potentially revealing deeper connections between chaos and order. The initial findings regarding the equilibrium's convergence suggest an intricate interplay that warrants further exploration, particularly in the context of physical systems.\\

Our conceptual approach to the 'distance to equilibrium' has shown promise in the analysis of biological signals, with forthcoming research aimed at solidifying its relevance and applicability. The exploration of entropy's various forms in conjunction with syntropy underscores the need for meticulous mathematical scrutiny to adequately articulate these relationships.\\

Finally, the application of vectorial transformations has unveiled the potential for uncovering latent information within systems, particularly when analyzing signals with embedded low-amplitude components or those affected by inadequate sampling rates. This insight opens new avenues for understanding the complexities inherent in signal analysis, highlighting the transformative power of syntropy and its associated functions in uncovering hidden dimensions of data.\\


\section*{Acknowledgments}

We extend our deepest gratitude to Dr. Edgardo Ugalde Saldaña, whose expertise in mathematical analysis provided invaluable insights that greatly enhanced this work. His guidance through the intricate mathematical concepts was instrumental in the development of the theories and applications presented in this article. His contributions are sincerely appreciated.

\section*{Annex}
The Java code provided by the author, which is used to calculate the equilibrium point \( p \) for any given system size \( N \), is listed below:

\begin{verbatim}
private static double syntropy(double p, double N) { 
  return (Math.exp(p) - 1) / (N * (Math.exp(1.0 / N) - 1)); 
}

private static double shannonEntropy(double p, double N){ 
  if (p == 0 || p == 1) return 0; 
  return -(p * Math.log(p) / Math.log(N) + (1 - p) * Math.log(1 - p) / Math.log(N)); 
}

public static double eqPoint(double N) {
  double lowerBound = 0.01; // Start slightly away from 0
  double upperBound = 0.99; // End slightly before 1
  double tolerance = 1e-16;
  double mid = 0;

  while ((upperBound - lowerBound) > tolerance) {
    mid = (lowerBound + upperBound) / 2;
    double midVal = syntropy(mid, N) - shannonEntropy(mid, N);
    if (midVal > 0){ upperBound = mid; } else { lowerBound = mid; }
  }

  return mid;
}
\end{verbatim}
\end{document}